# Effect of Mo doping for the Mn site in the ferromagnetic manganite $La_{0.7}Ca_{0.3}MnO_3$

G. Narsinga Rao and J. W. Chen*

**Department of Physics,
National Taiwan University
Taipei, Taiwan, R. O. C.**

**Abstract**

The structure, electronic, and magnetic properties of the Mo-doped perovskite $La_{0.7}Ca_{0.3}Mn_{1-x}Mo_xO_3$ ($x \leq 0.1$) have been studied. A significant increase in resistivity and lattice parameters were observed with Mo doping. A marginal decrease in the Curie temperature $T_c$ and the associated metal-insulator transition $T_p$ were observed. Magnetization data reveal that long-range ferromagnetic ordering persists in all samples studied and the saturation moment decreases linearly as x increases. Enhancement in magnetoresistance at near $T_c$ in the Mo-doped compounds with an optimum doping value x = 0.05 was observed. The overall experimental results can be explained by considering the induced $Mn^{2+}$ ions with $Mo^{6+}$ in the Mo-doped systems, with the strong FM coupling between $Mn^{4+/2+}$- O - $Mn^{3+}$.

*PACS:*   75.47. Gk; 75.47.Lx

*Corresponding author: Tel: 886-2-33665172; Fax: 886-2-23929578.

Email: jwchen@phys.ntu.edu.tw (J. W. Chen)

# 1. Introduction

Rare earth manganites with the general formula $R_{1-x}A_xMnO_3$ (R = rare-earth elements and A = alkali earth elements) exhibit wide variety of electronic and magnetic properties,[1,2] especially, the colossal magnetoresistance (CMR) effect near ferromagnetic transition temperature $T_c$ [3-5]. This phenomenon was attributed to the magnetic coupling between $Mn^{3+}$ and $Mn^{4+}$ ions described as Zener's double exchange (DE) [6,7] i.e. the hopping of $e_g$ electrons between the spin aligned $Mn^{3+}$ and $Mn^{4+}$ pairs. Recent studies [8,9] suggest that strong electron-phonon coupling arising from the deformation of the $MnO_6$ octahedra due to the Jahn-Teller effect also play a key role. The lattice distortion not only influences the effective transfer integral of the $e_g$ electrons but also the exchange interaction between manganese ions, as well as the magnetic structure of the compounds. Local distortion of the lattice around different ions could result in magnetic inhomogeneity in the compounds and give rise to an intriguing magnetic structure and magnetoresistance behavior. Since Mn-site substitution is a direct way to study the above-mentioned parameters related to the physical properties observed in the CMR systems, important clues concerning the mechanism of CMR can be obtained by studying the Mn-site doping effect.

Different effects can be produced upon Mn-site doping.[10-18] Except for the cases of Ru-[19,20] and Cr-doping [21,22], it has been found that doping in Mn sites with foreign elements in general destroys DE ferromagnetism in the samples thus decreases the Curie temperature and the metal insulator transition temperatures dramatically. The small values of decreasing rate $dT_c/dx$ with doping value x observed in the Cr (Ru-)doped samples were ascribed to the existence of ferromagnetic coupling between the $Cr^{3+}$-O-$Mn^{3+}$ and $Ru^{4+}$-O-$Mn^{3+}$. Recently, Raveau et al.[23,24] investigated effect of Mo-doping on the G-type antiferromagnetic $CaMn_{1-x}Mo_xO_3$ and found that ferromagnetic ordering occurs with $T_c \sim 125$ K in the doped samples with $x \leq 0.04$. These authors attributed such FM ordering to the DE coupling between $Mn^{3+}$ and $Mn^{4+}$ in the samples as $Mn^{3+}$ ions can be induced with hexavalent $Mo^{6+}$ doping. Since Mo can exist with different valence states, it is interesting to study the effect of Mo-doping on the CMR compound $La_{0.7}Ca_{0.3}MnO_3$. In this paper, we report the structure, electronic, and magnetic properties of the Mo-doped perovskite $La_{0.7}Ca_{0.3}Mn_{1-x}Mo_xO_3$ with $x \leq 0.1$. Our results indicate that the $Mn^{2+}$ ions with expense of $Mn^{4+}$ ions induced by $Mo^{6+}$ ions upon doping and the strong ferromagnetic coupling exists.

## 2. Experimental

Polycrystalline samples of nominal composition $La_{0.7}Ca_{0.3}Mn_{1-x}Mo_xO_3$ (x = 0, 0.025, 0.05, 0.075, and 0.1) were synthesized by solid state reaction method. Stoichiometric proportions of high purity $La_2O_3$ (99.999%), $CaCO_3$ (99.9%), $MnCO_3$ (99.99%), and $MoO_3$ (99.99%) powders were mixed and first fired in air at 900 °C for 12 h with an heating rate of 5 °C/min and cooled to room temperature. The pre-heated powders were well ground and calcined at 1200 °C for 24 h with several intermediate grindings. The calcined powders were then pressed into pellets of diameter 10 mm and sintered at 1400 °C for 24 h. The structure and phase purity of the samples were checked by powder X-ray diffraction (XRD) using the Cu-Kα radiation at room temperature. The field cooled (FC) and zero field cooled (ZFC) magnetization curves were measured in a commercial SQUID magnetometer with the temperature range from 4 K to 300 K in an applied field of 10 mT and the hysteresis curves were measured at 5 K for $-5\text{ T} \leq H \leq +5\text{ T}$. The dc resistivity and magnetoresistance (MR) of the samples were measured using the conventional four-probe method in the temperature range from 10 K to 350 K in a closed cycle helium refrigerator placed in a water cooled electromagnet with a field from -0.89 T to +0.89 T. The Curie temperature $T_c$ for the samples is defined as the inflection point of the M(T) curves at which the paramagnetic to ferromagnetic transition occurs in the sample. The MR ratio of the samples is defined as MR = $[\rho(H) - \rho(0)]/\rho(H)$, with $\rho(0)$ and $\rho(H)$ being the resistivity measured at zero magnetic field and an applied magnetic field H, respectively.

## 3. Results and discussion

The X-ray diffraction patterns of the polycrystalline $La_{0.7}Ca_{0.3}Mn_{1-x}Mo_xO_3$ samples are shown in figure 1(a). The XRD revealed single-phase patterns with no detected impurity peaks for all samples and the patterns can be indexed to orthorhombic perovskite structure with space group Pbnm. Fig. 1(b) gives the variation of the obtain lattice parameters with Mo doping content with no detected impurity peaks. As can be seen, the structure parameters increase monotonically with the increasing Mo- doping level $x$. The evolution of the lattice parameter with $x$ seems to suggest that the valence of Mo ion is 6+, because $Mo^{6+}$ (0.59 Å) ion has a large ionic radius than $Mn^{4+}$ (0.54 Å) ion[25].

To investigate more directly the valence state of Mo and Mn in $La_{0.7}Ca_{0.3}Mn_{1-x}Mo_xO_3$ samples, we carried XPS measurements on Mo 3d and Mn 2p regions. As shown in Figure 2 (a), the XPS spectra of Mo 3d exhibit a strong spin – orbit splitting with Mo $3d_{5/2}$ and $3d_{3/2}$ components for the sample x = 0.5. From this figure one can find that binding energies of Mo $3d_{5/2}$ and 3 $d_{3/2}$ are 233 and 236.3 eV respectively with a splitting of 3.3 eV. These observed binding energies are very close to the literature reported value of $Mo^{6+}$ ions[26]. Thus, the valence of Mo in our doped samples is 6+ (most stable valence of Mo in air, $d^0$ configuration). Figure 2(b) shows the spectra for Mn 2p level of the sample x = 0.05. The observed spin-orbit splitting between the two main peaks of the Mn $2p_{3/2}$ and Mn $2p_{1/2}$ states is ~ 12 eV[27] and the binding energy of peak position of Mn $2p_{3/2}$ level as 641.7 eV. From these observations, it is very difficult to estimate the valence state of manganese ions in our samples. The spectral features of the Mn 2p3/2 peaks in Fig. 2(b) are broad and asymmetric, which indicates the coexistence of $Mn^{3+}$ and $Mn^{4+}$ ions[28]. The asymmetric index β, defined as the ratio of half-width at half maximum on a high binding energy side, $W_H$, to half width at half maximum on a low binging energy side, $W_L$. The value of β was found to be 1.36, which is the evidence for the multiplet splitting of the Mn 2p level[28]. Thus, the manganese ions in our system have a highly mixed state. In addition, the core Mn 2p lines in figure 2(b) shows a satellite structure at higher binding energies, which is typical for $Mn^{2+}$ system such as MnO[29]. The higher oxidation state manganese ions do not show these high binding energy satellites[27,29]. From these spectra of Mn 2p level, we conclude that the mixed manganese ions $Mn^{2+}$, $Mn^{3+}$ and $Mn^{4+}$ exists in our doped samples..

Figure 3 shows the temperature dependence of field cooled (FC) and zero field cooled (ZFC) magnetization curves measured in an applied magnetic filed of 10 mT for the samples with x = 0, 0.05, and 0.10. The almost temperature-independent behavior observed in the ZFC M(T) curves indicates that long range FM ordering persists for all samples. A sharp transition from a paramagnetic (PM) to ferromagnetic (FM) state is observed for all samples. For the undoped sample, the obtained value of the Curie temperature $T_c$ is ~ 260 K. As shown in the inset of Fig. 3, the value of $T_c$ decreases monotonically with increasing Mo-doing x to ~ 210 K for the sample with x = 0.10. The decrease is fitted to a straight lines and the corresponding suppression rate, $dT_c/dx$ are found to be 2.4 K/at.% up to x = 0.5 and 6.2 K/at.% for higher Mo concentration. The decreasing rate $dT_c/dx$ observed here for the Mo-doped samples is much smaller than most of the other Mn-doped systems, where a much dramatic suppression of $T_c$ have been observed. This indicates that doping with Mo in $La_{0.7}Ca_{0.3}Mn_{1-x}Mo_xO_3$ does not merely destroy DE ferromagnetism since a much higher decreasing rate $dT_c/dx$ should be obtained. Instead, it indicates the existence of strong magnetic coupling between the neighboring Mn ions. Considering the charge neutrality, it is expected that the hexavalent Mo shift the average Mn valence. The partial substitution of hexavalent $Mo^{6+}$ not only decreases the $Mn^{4+}$ concentration, but it induces the $Mn^{2+}$ ions according to formula $La^{3+}_{0.7}Ca^{2+}_{0.3}Mn^{3+}_{0.7}Mn^{4+}_{0.3-2x}Mn^{2+}_{x}Mo^{6+}_{x}O_3$. The amount of the $Mn^{2+}$ will increase with increasing Mo content giving rise to the increase in $d_{Mn-O}$ and the decrease of $\theta_{Mn-O-Mn}$. Thus, the DE interaction becomes weakening because of the narrowing of the bandwidth and $T_c$ would decrease with the increase of the Mo-doping level. This formula is consistent with the XRD and XPS measurements. A smaller suppression rate and long-range ferromagnetic order in doped samples supports the opening of another DE channel between $Mn^{2+}$ - O - $Mn^{3+}$ ions.

On the other hand, suppose the $Mn^{2+}$ ions would not be introduced, the composition of $La_{0.7}Ca_{0.3}Mn_{1-x}Mo_xO_3$ ($0 \leq x \leq 0.10$) can be written as $La_{0.7}^{3+}Ca_{0.3}^{2+}Mn_{0.7+2x}^{3+}Mn_{0.3-3x}^{4+}Mo_x^{6+}O_3$. One can see that with the increase in x, the ratio of $Mn^{4+}/Mn^{3+}$ would decrease drastically and the $Mn^{4+}$ ion should disappear for x = 0.10. The observed FM ordering accompanied by I-M transition in doped samples is inconsistent with the obsence of $Mn^{2+}$. Based on the above consideration, we suggest that the mixed-valence state of $Mn^{2+}$, $Mn^{3+}$ and $Mn^{4+}$ co-exist in the doped samples.

Fig. 4 shows the temperature dependence of the inverse magnetic susceptibility $\chi$ for all samples. For a ferromagnet, it is well known that in the PM region, the relation between $\chi$ and the temperature T should follow the Curie–Weiss law, i.e. $\chi = C/(T-\Theta)$, where C is the Curie constant, and $\Theta$ is the Weiss temperature. The lines in Fig. 3(b) are the calculated curves deduced from the Curie–Weiss equation. Thus, the effective magnetic moment $\mu_{eff}$ can be obtained as 5.505, 5.894, 6.22, 6.025, and 6.01 $\mu_B$ for the samples with x = 0, 0.025, 0.05, 0.075, and 0.10, respectively. According to a mean field approximation, the expected $\mu_{eff}$ can also be calculated as 4.59, 4.545, 4.499, 4.453, and 4.407 $\mu_B$ for the sample with x = 0, 0.025, 0.050, 0.075, and 0.10, respectively. These calculated values are smaller than the fitted values from $1/\chi$ vs T curves. It implies that the short-range ferromagnetic interaction exist above transition temperature[30]. The observed enhancement of $\mu_{eff}$ with Mo doping also supports the existence of $Mn^{2+}$.

Fig. 5 shows the typical isothermal M(H) curves of magnetization for x = 0, 0.05 and 0.10 samples measured at 5 K for -5 T $\leq$ H $\leq$ 5 T. Saturation of the M(H) curves at low fields again reveals the existence of long-range ferromagnetic order in all samples. The saturated magnetic moment $\mu_{exp}$ per formula unit for various value of x can be deduced from the experimental saturation magnetization $M_s$. The results are shown in the inset of figure 5. The obtained value for the undoped sample is 3.65 $\mu_B$/f.u., which is close to the value (3.7 $\mu_B$/f.u.) that DE model predicts. Upon doping, $\mu_B$ decreases linearly with x following the relation $\mu_B(x) = (3.65 - 3.15 x) \mu_B$ to a value of 3.33 $\mu_B$/f.u. for x = 0.10.

The temperature dependence of resistivity $\rho(T)$ curves for the $La_{0.7}Ca_{0.3}Mn_{1-x}Mo_xO_3$ samples taken at H = 0 and in an applied field H = 0.8 T are plotted in Fig. 6 (a). For H = 0, the $\rho(T)$ curve depicts a single sharp metal-insulator transition near the Curie temperature $T_c$ for the undoped sample. With increasing Mo content x, the metal insulator transition temperature $T_p$ shifts to lower temperature with suppression rate similar to $T_c$, while the room temperature resistivity $\rho_{RT}$ increases (Fig. 6 (b)). Applying the magnetic field H lowers the resistivity for all the samples and leads to an increase of $T_p$ and a negative CMR effect. The value of resistivity increase monotonously with the increase of Mo-doping level in the whole temperature range studied. When partial $Mn^{4+}$ ions was replaced by $Mo^{6+}$ ions, correspondingly, some other $Mo^{4+}$ ions should convert to $Mn^{2+}$ ions in order to keep the balance of the valence in formula. In other words, the delocalized number of electrons of Mn ions decreased. At the same time, the random distribution of Mo ions at Mn sublattice will introduce the appearance of random potential around Mo ions, which can localize electrons of Mn ions and result in the rapid increase of resistivity. Under the impact of these two elements, the resisitivity increase monotonously with the increase of Mo-doping content.

The temperature dependence of MR ratio measured at H = 0.8 T for the $La_{0.7}Ca_{0.3}Mn_{1-x}Mo_xO_3$ system is plotted in figure 7 (a).  The MR versus T curves reveal typical feature of the intrinsic CMR behavior and exhibit a very sharp peak near the M-I transition temperature $T_p$ for all samples.  Enhancement of the MR ratio is observed in the Mo-doped samples with x ≤ 0.05.  For samples with x > 0.05, the MR ratio begins to decreases with increasing Mo-doping concentration x.  For x = 0.05, MR reaches a maximum value of 110 % at $T_p$, which is double than that of undoped sample for H = 0.8 T.  The existence of optimum doping level that favors CMR for x = 0.05 is evident, as can be seen from Fig. 7 (b), where the MR ratio at $T_p$ are plotted as a function of the doping concentration x.

Substitution of Mn site with paramagnetic unfilled $e_g$ orbital (Ru, Cr, Co, Cu, Fe, and Ni) ions [14-22] in the CMR manganites in general causes a phase separation in the doped samples.  Such phase separation will lead to cluster or spin glass type behavior in the low temperature as revealed from the magnetization data. In addition, an additional broad hump in the resistivity curve was usually observed irrespective of the ground state of the parent compound. In contrast, in the present study the generic feature of additional hump in the resistivity curve and cluster or spin glass type behavior is absent in Mo-doped samples. This indicates that there is no phase separation in $La_{0.7}Ca_{0.3}Mn_{1-x}Mo_xO_3$.  The almost coincidence of the Curie temperature and the M-I transition temperature in these samples reveals that the occupation inhomogeneity is negligible.   The observed sharp peaks in the MR ratio near $T_p$ for all samples reconfirm above conjectures. To our knowledge, this is the first existing system that shows no phase separation and inhomogeneity upon Mn-site doping.

## 4. Conclusion

We have investigated the effects of Mo doping in $La_{0.7}Ca_{0.3}MnO_3$ through magnetic and transport measurements. Both M(T) and $\rho$(T) data reveal no phase separation or inhomogeneity in the Mo-doped samples. Enhancement in the MR ratio near $T_p$ with an optimum doping value x = 0.05 was observed. In addition, the saturation moment was found to decrease linearly with increasing Mo-doping x. The results indicate that the $Mn^{2+}$ ions induced by $Mo^{6+}$ ions in the $La_{0.7}Ca_{0.3}Mn_{1-x}Mo_xO_3$ compounds. The existence of FM coupling between $Mn^{4+}$-O-$Mn^{3+}$ and $Mn^{2+}$-O-$Mn^{3+}$ in the samples is responsible for slow decreasing rate, $dT_c/dx \sim 2.4$ K/at.%, observed in the systems.

**Acknowledgements**

  This work was supported by the National Science Council of ROC under grant No. NSC95-2112-M-002-049-MY3.

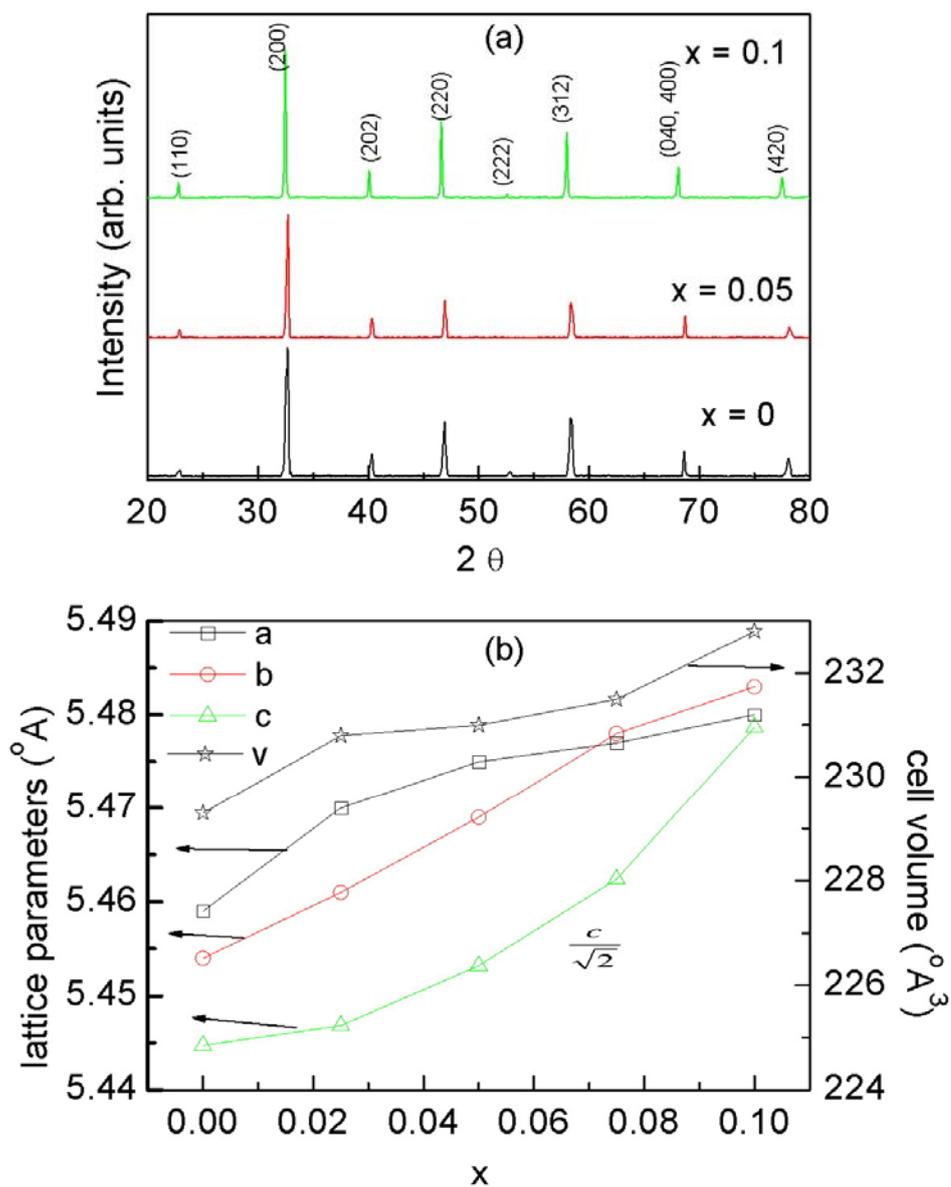

Figure 1 (a). The X-ray diffraction patterns of the polycrystalline $La_{0.7}Ca_{0.3}Mn_{1-x}Mo_xO_3$ samples, (b). The variation of lattice parameters and unit cell volume as a function of x.

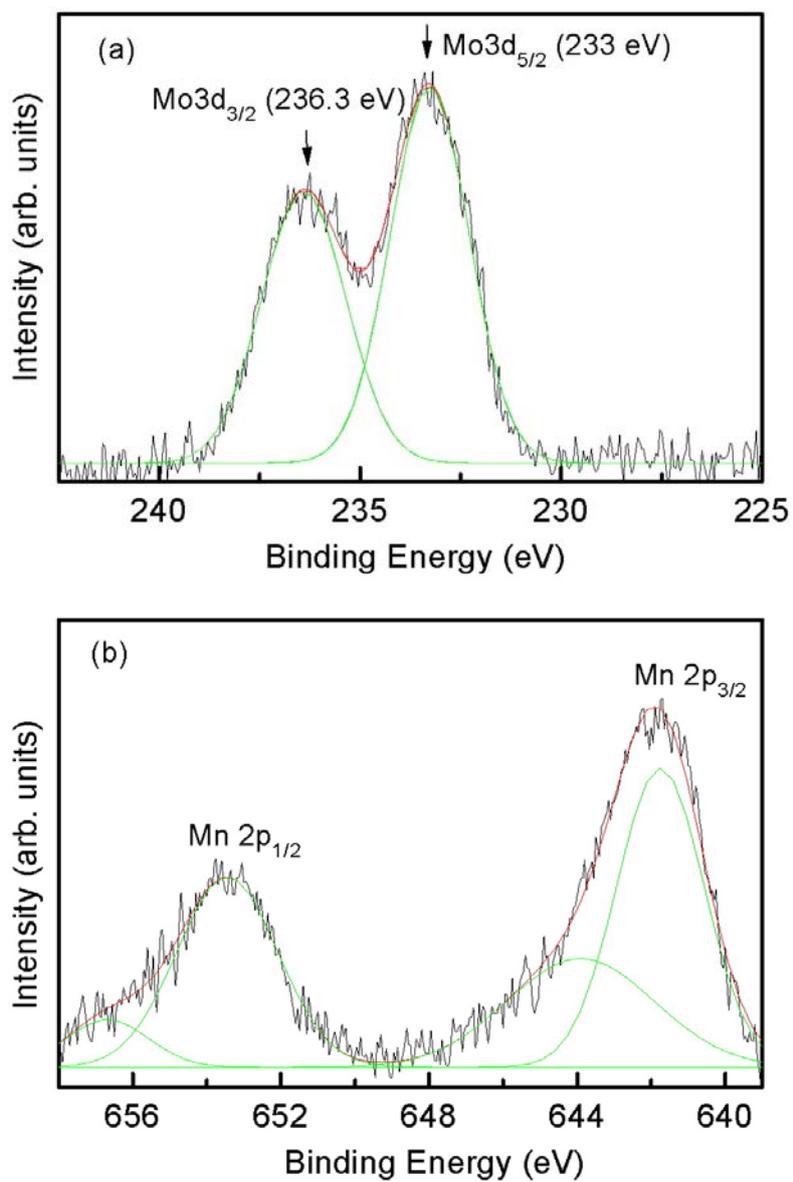

Figure 2 (a) The XPS spectra of (a) Mo 3d and (b) Mn 2p regions for the x = 0.05 sample.

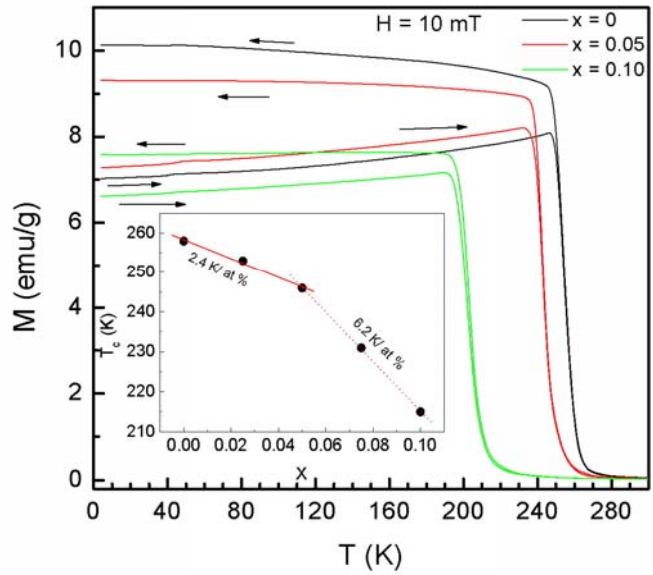

Figure 3. The temperature dependence of field cooled (FC) and zero field cooled (ZFC) magnetization curves measured in an applied magnetic filed of 10 mT for x = 0, 0.05 and 0.10 samples. The inset shows the Curie temperature $T_c$ as a function of *x*.

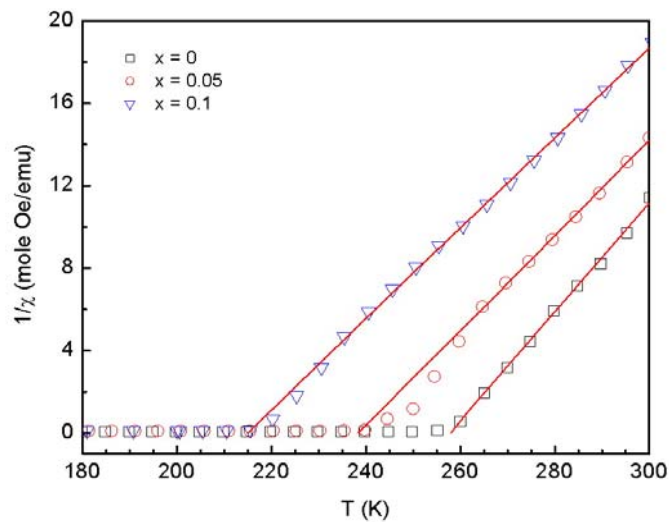

Figure 4. Inverse of magnetic susceptibility as a function of temperature for the samples with x = 0, 0.5, and 0.10. The lines are fitted curves according to the Curie-Weiss law.

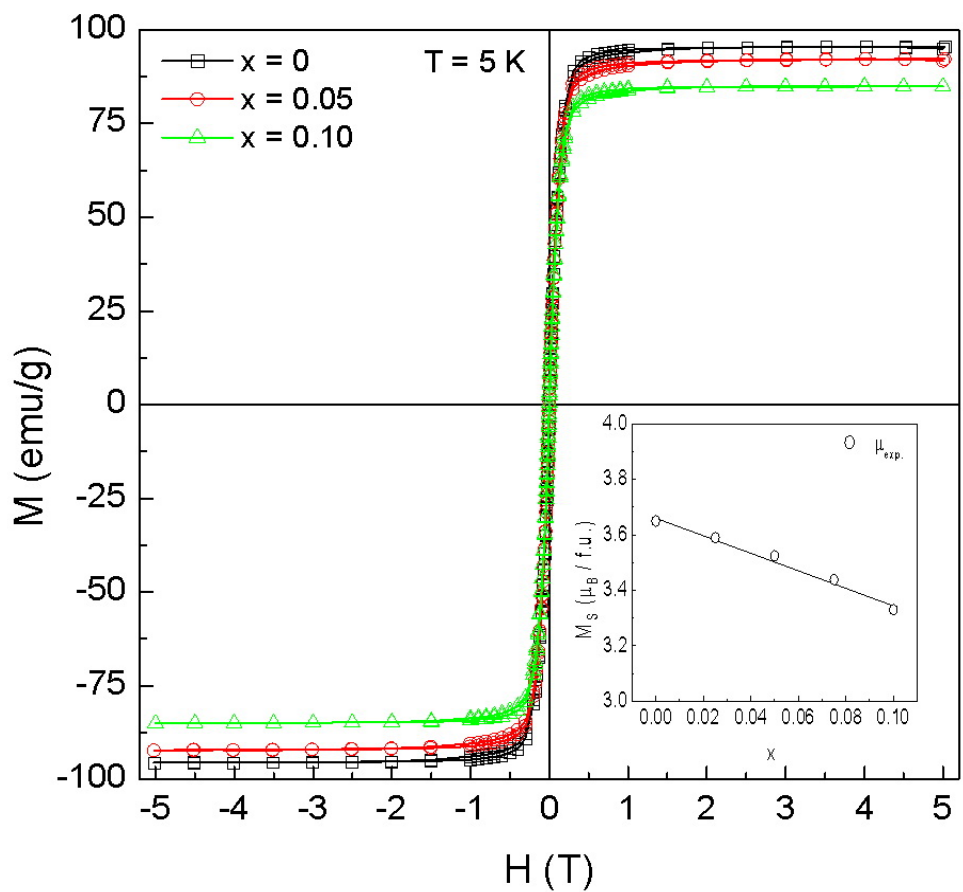

Figure 5. The isothermal M (H) curves for x = 0, 0.05, and 0.10 samples measured at 5 K for -5 T ≤ H ≤ 5 T. The inset shows the experimental saturated magnetic moment per formula unit as a function of *x* at 5 K.

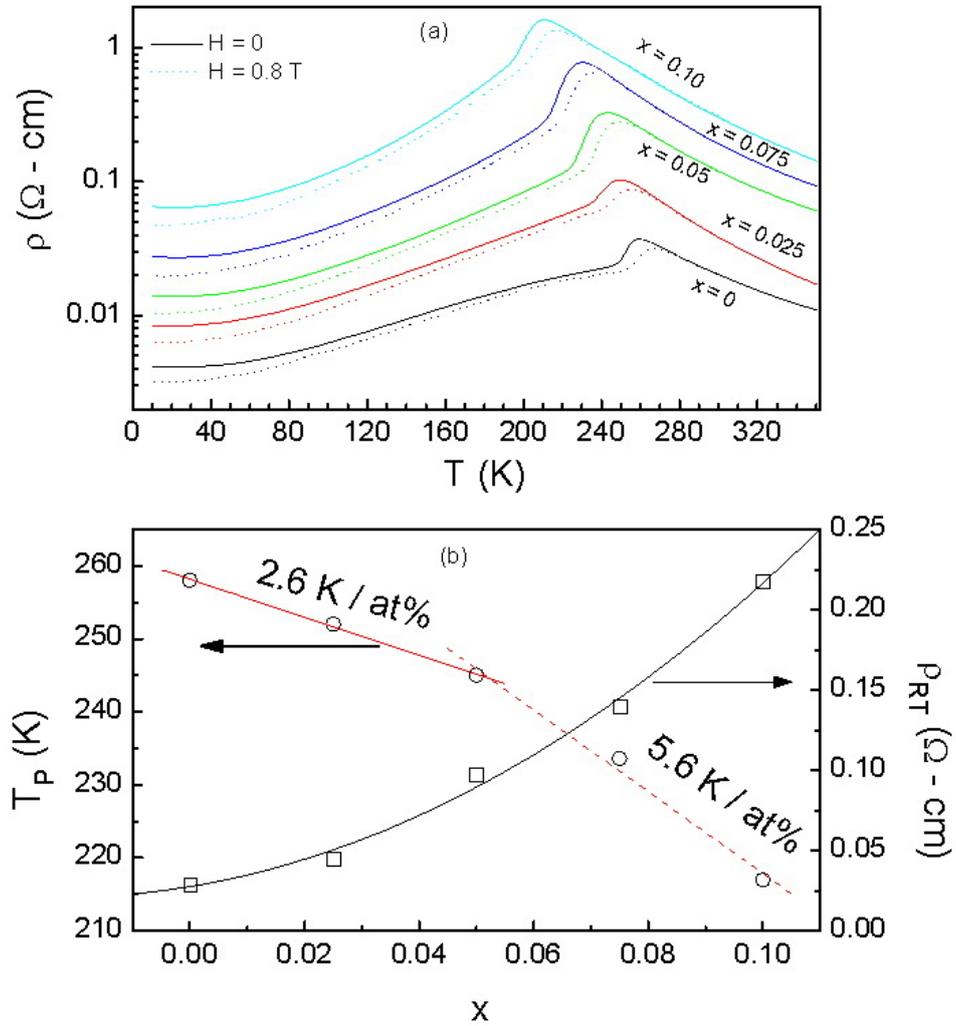

Figure 6 (a). The temperature dependence of resistivity ρ (T) curves for the La$_{0.7}$Ca$_{0.3}$Mn$_{1-x}$Mo$_x$O$_3$ samples taken at H = 0 and in an applied field H = 0.8 T, (b). The metal insulator transition temperature T$_p$ and room temperature resistivity as a function of x.

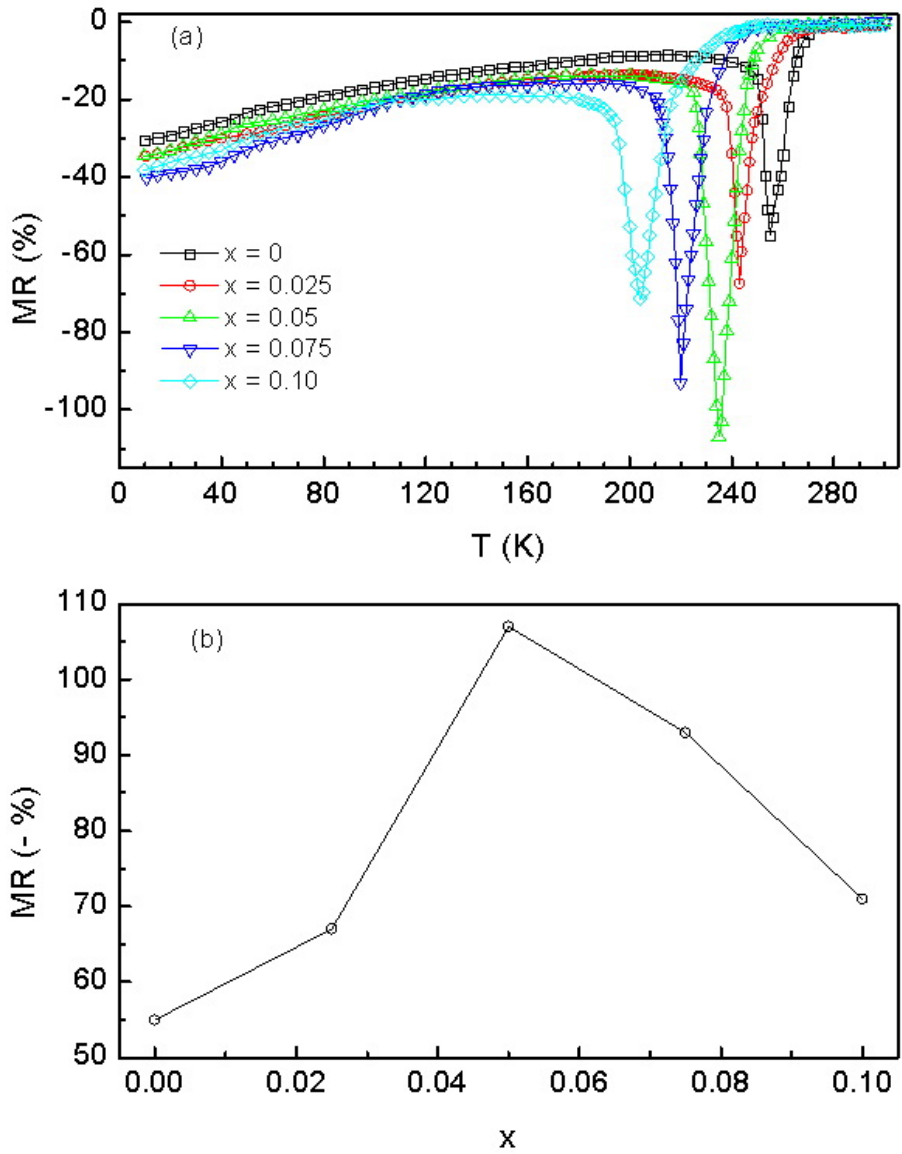

Figure 7 (a). The temperature dependence of MR ratio measured at H = 0.8 T for the La$_{0.7}$Ca$_{0.3}$Mn$_{1-x}$Mo$_x$O$_3$ samples, (b). The MR ratio at T$_p$ are plotted as a function of the doping concentration x.